\documentclass{ifacconf}
\usepackage{natbib}   
\usepackage{graphicx}
\usepackage{textcomp}
\usepackage{microtype}
\usepackage{amsmath,amssymb,bm,bbm,mathrsfs,amscd}
\usepackage{calc}
\usepackage{xcolor}
\usepackage{tikz}
\usetikzlibrary{arrows.meta, positioning}
\usepackage{dsfont}
\usepackage{epstopdf}
\usepackage{amsfonts}
\usepackage{rotating}
\usepackage{mathtools}
\usepackage{subfig}
\usepackage{enumerate,pgfplots}
\pgfplotsset{compat=1.18}

\newtheorem{assumption}{Assumption}
\newcommand{\ba}{\begin{array}}
\newcommand{\ea}{\end{array}}

\newcommand{\be}{\begin{equation}}
\newcommand{\ee}{\end{equation}}

\newcommand{\mc}{\mathcal}

\newcommand{\R}{\mathbb{R}}

\newcommand{\V}{\mathcal{V}}

\newcommand{\G}{\mathcal{G}}

\def\R{\mathbb{R}}

\def\diag{{\rm diag}\,}
\begin{document}
\begin{frontmatter}
\title{On a Coupled Adoption-Opinion Framework for Competing Innovations} 

\thanks[footnoteinfo]{This work was supported in part by the Wallenberg AI, Autonomous Systems and Software Program (WASP) funded by the Knut and Alice Wallenberg Foundation, and by the European Union -- Next Generation EU, Mission 4, Component 1, under the PRIN project {\em{TECHIE: A control and network-based approach for fostering the adoption of new technologies in the ecological transition}}, Cod. 2022KPHA24, CUP Master: D53D23001320006, CUP: B53D23002760006.}

\author[KTH]{Martina Alutto} 
\author[Polito]{Fabrizio Dabbene} 
\author[KTH]{Angela Fontan}
\author[KTH]{Karl H. Johansson}
\author[Polito]{Chiara Ravazzi} 

\address[KTH]{Division of Decision and Control Systems, School of Electrical Engineering and Computer Science, \\KTH Royal Institute of Technology, Stockholm, Sweden \\(e-mails: \{alutto; angfon; kallej\}@kth.se)}
\address[Polito]{Institute of Electronics, Computer and Telecommunication Engineering, National Research Council of Italy, Politecnico di Torino, 10129 Torino, Italy (e-mails: \{chiararavazzi;\,fabriziodabbene\}@cnr.it)}

\begin{abstract}
    In this paper, we propose a two-layer adoption-opinion model to study the diffusion of two competing technologies within a population whose opinions evolve under social influence and adoption-driven feedback. After adopting one technology, individuals may become dissatisfied and switch to the alternative. We prove the existence and uniqueness of the adoption-diffused equilibrium, showing that both technologies coexist and that neither partial-adoption nor monopoly can arise. Numerical simulations show that while opinions shape the equilibrium adoption levels, the relative market share between the two technologies depends solely on their user-experience. As a consequence, interventions that symmetrically boost opinions or adoption can disproportionately favor the higher-quality technology, illustrating how symmetric control actions may generate asymmetric outcomes. 
\end{abstract}


\end{frontmatter}

\thinmuskip=0.5mu  
\medmuskip=0.5mu   
\thickmuskip=0.5mu 

\section{Introduction}
Innovation diffusion processes, describing the spread of new ideas, technologies, behaviors or practices within a social system, have been extensively investigated across different disciplines \citep{Rogers2010}. 
Numerous studies examine how individuals adopt innovations under the influence of both external exposure and peer influence \citep{BRESCHI2023103651, VILLA2024106106}.  
Classical economic approaches, such as the Bass model \citep{Bass1969}, formalize adoption as the combined effect of independent decisions and imitation. Threshold-based models show that widespread adoption can occur once a critical mass of adopters is reached \citep{Granovetter1978}. More generally, network-based frameworks, such as the linear threshold and independent cascade models \citep{Kempe2003}, describe the spread of innovation as a contagion process, in which individuals change their behavior when the influence of their social contacts exceeds a tolerance threshold. Overall, these perspectives highlight that the diffusion of innovation is a form of complex contagion, characterized by nonlinear adoption responses and strongly influenced by both social dynamics and external factors \citep{centola2018behavior}.

Epidemic-inspired compartmental models have emerged as an established paradigm to study diffusion dynamics. Initially developed for infectious diseases \citep{Kermack.McKendrick:1927}, SIR and SIS models have been adapted to represent the propagation of behaviors, technologies, and information \citep{
goffman1966mathematical, bettencourt2006power, Pastor-Satorras2015}. Variants including mechanisms of reversible adoption or dissatisfaction, such as the SIRS framework \citep{LI20141042, ZHANG2021126524}, provide a richer representation. 

It is worth noting that adoption is also strongly influenced by beliefs, attitudes, and perceived social norms. Opinion dynamics models provide a complementary perspective to capture these effects. The DeGroot model \citep{DeGroot1974} explains consensus formation through iterative averaging of neighbors’ views, whereas the Altafini’s model \citep{Altafini2012} allows for antagonistic relationships and persistent disagreement. The Friedkin–Johnsen model \citep{Friedkin1990} further incorporates individual predispositions, highlighting how resistance and heterogeneity naturally emerge in social decision processes. Building on this, a growing body of work integrates epidemic-like adoption dynamics with opinion evolution, acknowledging their mutual reinforcement: opinions shape willingness to adopt, while adoption feeds back into beliefs \citep{granell2013dynamical, wang2019impact, lin2021discrete, she2021peak, bizyaeva2024active, Xu2024}. 
These coupled frameworks reveal complex outcomes that traditional contagion-only models cannot capture.
Recent contributions also address how to effectively intervene in such systems. \cite{alutto2025predictive} propose a predictive control framework where different policies, shaping opinions, adoption enhancement, and dissatisfaction reduction, are compared, and show that adaptive strategies outperform the static ones. Then, \cite{alutto2025modeling} extend this modeling to data-driven heterogeneous populations, embedding demographic factors such as mobility and age to better reflect real-world variation in influence and propensity to adopt.

However, in many realistic scenarios, individuals face multiple competing technologies, making adoption a strategic decision influenced by peer coordination, as in the case of mobile service providers, car-sharing services, or streaming platforms. 
To capture this strategic dimension, game-theoretic models based on network coordination games have gained increasing attention \citep{montanari2010spread,young2011dynamics}, formalizing technology adoption as a process in which individuals select the option that maximizes expected utility given their neighbors’ choices. 
Theoretical studies have shown that a variety of outcomes may arise from such competitive dynamics. While traditional economic theory suggests convergence to a single dominant technology \citep{david1985clio,arthur1989competing,witt1997lock}, more recent works reveal that monopolistic dominance is not the only possible scenario: heterogeneity, segmentation, switching costs, and asymmetric network effects can enable long-term coexistence. Moreover, allowing dual users in complex contagion processes further broadens the parameter region supporting coexistence \citep{min2018competing}. Similar phenomena are also observed in epidemic models of multi-virus or multi-contagion spreading \citep{liu2019analysis,gracy2025modeling}.

Existing models often treat competition and opinion dynamics separately. To address this gap, we propose a novel adoption-opinion model to study the diffusion of two competing technologies within a population structured as a two-layer network. Adoption evolves through epidemic-inspired compartmental dynamics with susceptible, adopter, and dissatisfied compartments for both technologies, while opinions follow a modified Friedkin–Johnsen process, integrating social influence and feedback from adoption levels. Analytical results establish the existence and uniqueness of an adoption-diffused equilibrium point, showing that at equilibrium both technologies coexist and no partial-adoption or monopoly states are possible.

The rest of the paper is organized as follows. Section~\ref{sec:model} introduces the adoption-opinion model. Stability results are provided in Section \ref{sec:stability}, while Section~\ref{sec:simulations} presents numerical simulations. Finally, Section~\ref{sec:conclusion} concludes the paper and outlines directions for future research.

\subsection{Notation}
We denote by $\R$ and $\R_{+}$ the sets of real and nonnegative real numbers, respectively, while $\R_{+}^{n \times n}$ indicates the set of real matrices with dimension $n \times n$ and nonnegative entries. 
The all-1 vector and the all-0 vector are denoted by $\boldsymbol{1}$ and $\boldsymbol{0}$, respectively. The identity matrix and the all-0 matrix are denoted by $I$ and $\mathbb{O}$, respectively.
The transpose of a matrix $A$ is denoted by $A^T$. 
For $x$ in $\R^n$, let $||x||_1=\sum_i|x_i|$ and $||x||_{\infty}=\max_i|x_i|$ be its $l_1$ and $l_{\infty}$ norms, while $\diag(x)$ denotes the diagonal matrix whose diagonal coincides with $x$. 
For an irreducible matrix $A$ in $\R_+^{n\times n}$, i.e., a matrix whose associated directed graph is strongly connected, we let $\rho(A)$ denote the spectral radius of $A$.  
Inequalities between two vectors $x$ and $y$ in $\R^n$ are meant to hold true entry-wise, i.e., $x \le y$ means that $x_i\le y_i$ for every $i$, $x< y$ means that $x_i< y_i$ for every $i$, and $x\lneq y$ means that $x_i\le  y_i$ for every $i$ and $x_j<y_j$ for some $j$; analogous holds for matrices. 



\section{Model Description}\label{sec:model}
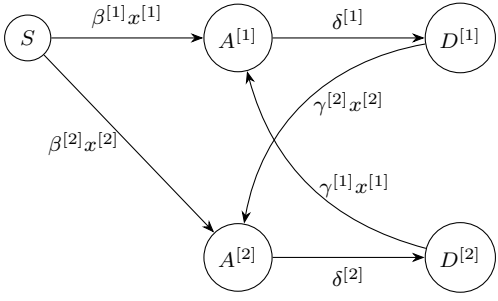
\begin{figure}
    \centering
    \begin{tikzpicture}[node distance=2cm, every node/.style={font=\small}, >=Stealth]
        \node[draw, circle] (S) {$S$};
        \node[draw, circle, right=of S] (A1) {$A^{[1]}$};
        \node[draw, circle, right=of A1] (D1) {$D^{[1]}$};
        \node[draw, circle, right=of S, below=of A1] (A2) {$A^{[2]}$};
        \node[draw, circle, right=of A2] (D2) {$D^{[2]}$};

        \draw[->] (S) -- node[above] {$\beta^{[1]} x^{[1]}$} (A1);
        \draw[->] (A1) -- node[above] {$\delta^{[1]}$} (D1);
        \draw[->] (D1) edge[bend right=35, below] node[right] {$\gamma^{[2]} x^{[2]}$} (A2);
        \draw[->] (S) -- node[left] {$\beta^{[2]} x^{[2]}$} (A2);
        \draw[->] (A2) -- node[below] {$\delta^{[2]}$} (D2);
        \draw[->] (D2) edge[bend left=30, above] node[right] {$\gamma^{[1]} x^{[1]}$} (A1);

    \end{tikzpicture}
    \caption{Coupled transition diagram for two competing technologies. Susceptible individuals can adopt either technology, moving to adoption compartments $A^{[1]}$ or $A^{[2]}$. Adopters may become dissatisfied, entering $D^{[1]}$ or $D^{[2]}$. Dissatisfied individuals can switch to the other technology at rates modulated by the opinion variables $x^{[k]}$.}
    \label{fig:transitions2}
\end{figure}

We consider a population divided into $n$ communities, where each community represents a subpopulation of indistinguishable individuals that can interact and adopt two competing technologies, denoted by $1$ and $2$. Within each community $i \in \mathcal{V}= \{ 1,\dots,n\}$, individuals are classified into the following compartments at each time instant $t\geq0$:
\begin{itemize}
    \item $s_i(t)$: the fraction of susceptibles to both technologies,
    \item $a_i^{[1]}(t)$, $a_i^{[2]}(t)$: the fraction of adopters of technology $1$ and $2$, respectively,
    \item $d_i^{[1]}(t)$, $d_i^{[2]}(t)$: the fraction of dissatisfied of technology $1$ and $2$, respectively.
\end{itemize}
The interactions among communities are captured by a weighted directed network $\mc G = (\mc V, \mc E, W)$, where $\mc V $ is the set of nodes, $\mc E \subseteq \mc V \times \mc V$ is the set of directed links and $W$ is a nonnegative weight matrix, referred to as \emph{physical interaction matrix}.
Each entry $W_{ij}$ quantifies the strength of influence that community $j$ exerts on community $i$ in terms of adoption exposure. The three-state adoption process is illustrated in Figure~\ref{fig:transitions2}, where arrows indicate the possible transitions between the susceptible ($S$), adopter ($A$), and dissatisfied ($D$) compartments. Susceptible individuals can adopt either technology, entering the corresponding adoption compartment $A^{[k]}$. From the adoption state, individuals may become dissatisfied at a technology-specific rate, moving to the dissatisfied compartments $D^{[k]}$. Dissatisfied individuals of one technology can switch to the other technology if their opinion favors it.
Indeed, each community $i$ is also associated with an \emph{opinion variable} $x_i^{[k]} \in [0,1]$ reflecting the average interest toward adopting technology $k$. Opinions evolve over a social network $\mc{\tilde G} = (\mc V, \mc{\tilde E}, \mc{\tilde{W}})$, where $\tilde{\mc E} \subseteq \mc V \times \mc V$ is the set of directed links, and $\tilde{W} \in \R_+^{n \times n}$ is a nonnegative \emph{social interaction matrix}.

Before introducing the full dynamics, we emphasize that the adoption process is influenced by opinions. In particular, the susceptibility of an individual to adopt a technology, either from the susceptible state or from dissatisfaction with the other technology, depends on the opinion variable $x_i^{[k]}$ toward this technology. These adoption rates are modeled by the functions 
$\beta_i^{[k]} : [0,1] \rightarrow [0,1]$ and $\gamma_i^{[k]} : (0,1] \rightarrow [0,1]$. We assume a simple linear dependence on opinions: $\beta_i^{[k]}(x_i^{[k]}) = \beta_i^{[k]}\, x_i^{[k]}$ and $\gamma_i^{[k]}(x_i^{[k]})= \gamma_i^{[k]}\, x_i^{[k]}$ with $\beta_i^{[k]} \in [0,1]$ and $\gamma_i^{[k]} \in (0,1)$ for each technology $k \in \{1,2\}$. We denote by $\delta_i^{[k]} \in [0,1]$ the rate at which individuals become dissatisfied of technology $k$. Unlike the adoption rates, the dissatisfaction rate is assumed independent of the opinion variables, which could reflect the tendency to abandon a technology due to the experienced quality of the service or technology (e.g., user experience, reliability, performance), rather than being driven by social influence or prior inclination.
Formally, the \emph{coupled discrete-time adoption-opinion dynamics} is given by
\begin{subequations}\label{eq:model}
    \begin{align}
s_i(t+1)=& s_i(t) - s_i(t)\sum_{k'\in \{1,2\}} \beta_i^{[k']} x_i^{[k']}(t) \sum_{j=1}^n W_{ij} a_j^{[k']}(t)  \label{eq:dyn_s}\\
a_i^{[k]}(t+1) =& a_i^{[k]}(t) + \beta_i^{[k]} x_i^{[k]}(t) s_i(t) \sum_{j=1}^n W_{ij} a_j^{[k]}(t)   \nonumber\\
    &\quad - \delta_i^{[k]}\, a_i^{[k]}(t) + \gamma_i^{[k]}\, x_i^{[k]}(t)d_i^{[\ell]}(t), \label{eq:dyn_a} \\
d_i^{[k]}(t+1) =& d_i^{[k]}(t) - \gamma_i^{[\ell]}\, x_i^{[\ell]}(t)\, d_i^{[k]}(t)
    + \delta_i^{[k]}\, a_i^{[k]}(t), \label{eq:dyn_d} \\
x_i^{[k]}(t+1) =& (1 - \lambda_i^{[k]} - \xi_i^{[k]}) x_i^{[k]}(0)
    + \lambda_i^{[k]} \sum_{j=1}^n \tilde W_{ij} x_j^{[k]}(t)\nonumber\\
    &\quad + \xi_i^{[k]} \sum_{j=1}^n W_{ij} a_j^{[k]}(t), \label{eq:dyn_x} 
\end{align}
\end{subequations}
for each node $i \in \mc V$, $k,\ell \in \{1,2\}$, and $\ell \neq k$.
    
The opinion dynamics \eqref{eq:dyn_x} is a modified version of the Friedkin--Johnsen framework \citep{Friedkin1990}; $x_i(0)$ represents the initial predisposition of community $i$ towards technology $k$, and the coefficients $\lambda_i^{[k]},\, \xi_i^{[k]} \,\, \geq 0$ satisfy $\lambda_i^{[k]} + \xi_i^{[k]} < 1$ for all $i$ and for each technology $k$. These parameters weigh three different sources of influence: intrinsic predisposition, social neighborhood, and the observed adoption behavior in the physical layer for each technology. A larger $\lambda^{[k]}_i$ makes opinions more sensitive to neighbors’ beliefs, whereas higher $\xi_i$ emphasizes adoption-driven feedback.

The next assumption guarantees that the physical network is connected and that, in the social network, each community is directly or indirectly influenced by at least one stubborn community for each technology $k$.
\begin{assumption}\label{ass:ass1}
    Both $W$ and $\tilde{W}$ are row-stochastic (i.e., $\sum_j W_{ij} =\sum_j \tilde W_{ij} = 1$, for all $i$). The matrix $W$ is irreducible. For any technology $k$ and any node $i\in\V$, it holds that $\xi_i^{[k]} >0$ and there exists a path in $\widetilde{\G}$ from $i$ to $j$ with $\lambda_j^{[k]}<1$ and $x_j^{[k]}(0)>0$. In addition, it holds that $\beta^{[1]}_i + \beta^{[2]}_i \in (0,1)$ for all $i \in\mc V$.
\end{assumption}\medskip 

The system in \eqref{eq:model} can be rewritten compactly as:  
{\thinmuskip=0mu  
\medmuskip=0mu   
\thickmuskip=0mu 
\begin{subequations}\label{eq:vector_model} \begin{align} 
\mspace{-5mu} s(t+1) =& s(t) - \sum_{k'} \diag(s(t)) B^{[k']} X^{[k']}(t)  W a^{[k']}(t), \\[1ex] 
\mspace{-5mu} a^{[k]}(t+1) =& a^{[k]}(t) - \Delta^{[k]} a^{[k]}(t)
    + \Gamma^{[k]} X^{[k]}(t) d^{[\ell]}(t) \nonumber\\
    &\mspace{5mu}+ B^{[k]} X^{[k]}(t) \diag(s(t)) W a^{[k]}(t), \\[1ex] 
\mspace{-5mu}d^{[k]}(t+1) =& d^{[k]}(t) - \Gamma^{[\ell]} X^{[\ell]}(t) d^{[k]}(t)
    + \Delta^{[k]} a^{[k]}(t), \\[1ex]
\mspace{-5mu}x^{[k]}(t+1) =&(I - \Lambda^{[k]} - \Xi^{[k]}) x^{[k]}(0)
    + \Lambda^{[k]} \tilde W x^{[k]}(t)
    + \Xi^{[k]} W a^{[k]}(t),
\end{align}\end{subequations}}
for $k,\ell \in \{1,2\}$ and $\ell \neq k$, where $X^{[k]}(t):=\diag(x^{[k]}(t))$, $\Delta^{[k]} :=  \mathrm{diag}(\delta^{[k]})$, $B^{[k]} :=  \mathrm{diag}(\beta^{[k]})$, $\Gamma^{[k]} :=  \mathrm{diag}(\gamma^{[k]})$, $\Lambda^{[k]}:=\mathrm{diag}(\lambda^{[k]})$, and $\Xi^{[k]}:=\mathrm{diag}(\xi^{[k]})$.

For simplicity of notation, from now on, the state of the adoption model \eqref{eq:vector_model} at time $t\geq0$ is denoted as 
$$y(t):= (s(t), a^{[1]}(t),a^{[2]}(t),d^{[1]}(t), d^{[2]}(t),x^{[1]}(t), x^{[2]}(t) ).$$ 
Note that $y(t)$ is in $\mathbb{R}^{7n}$.

\begin{prop}\label{prop:invariant}
    Consider the adoption-opinion model \eqref{eq:vector_model} under Assumption \ref{ass:ass1}. 
    If the initial condition $y(0)$ is in $[0,1]^{7n}$ and 
    \be \label{eq:norm} s(0) + \sum_k (a^{[k]}(0) + d^{[k]}(0)) = \boldsymbol{1},\ee
    then for all $t \ge 0$ it holds that $y(t)$ is in $[0,1]^{7n}$ and 
    \be \label{eq:norm-t} s(t) + \sum_k (a^{[k]}(t) + d^{[k]}(t)) = \boldsymbol{1}.\ee
\end{prop}
\begin{pf}
    We show the well-posedness of the adoption-opinion model in \eqref{eq:vector_model} by induction. Suppose that a time $t \geq0$, $y(t)$ is in $[0,1]^{7n}$ with $ s(t) + \sum_k (a^{[k]}(t) + d^{[k]}(t)) = \boldsymbol{1}$. From \eqref{eq:dyn_s}-\eqref{eq:dyn_d}, for each node $i$ it holds that:
    \begin{align*}
        &s_i(t+1) = s_i(t) \Big[ 1- \sum_{k'} \beta_i^{[k']}\, x_i^{[k']}(t)\,\sum_{j=1}^n W_{ij} a_j^{[k']}(t) \Big] \geq 0,\\
        &a^{[k]}_i(t+1) \geq  a^{[k]}_i(t) (1-\delta^{[k]}_i) \geq 0, \quad \forall k\in\{1,2\}\\
        &d^{[k]}_i(t+1) \geq  d^{[k]}_i(t) (1-\gamma^{[\ell]}_i x^{[\ell]}_i(t))\geq 0, \quad \forall k,l \in\{1,2\},\,\, l\neq k,
    \end{align*}
    where the first inequality follows also from Assumption \ref{ass:ass1}, while the second and third inequality follow from the definition of parameters $\delta^{[k]}_i$ and $\gamma_i^{[k]}$. 
    Moreover, 
    \begin{align*}
        s(t+1) + \sum_k (a^{[k]}(t+1) + d^{[k]}(t+1)) =\\
        = s(t) + \sum_k (a^{[k]}(t) + d^{[k]}(t)) = \boldsymbol{1}.
    \end{align*}
    Note also that, from the opinion dynamics in \eqref{eq:dyn_x}, we get  
    \begin{align}
        x^{[k]}(t+1) &\geq (I-\Lambda^{[k]}-\Xi^{[k]})x(0) \geq \boldsymbol{0}, \label{eq:x-inf}\\
        x^{[k]}(t+1) &\leq (I-\Lambda^{[k]}-\Xi^{[k]})x(0)+ \Lambda^{[k]} \boldsymbol{1} + \Xi^{[k]} \boldsymbol{1} \leq \boldsymbol{1},
    \end{align}
    using the fact that $x^{[k]}(t), a^{[k]}(t) \in [0,1]^n$, the row-stochasticity of $W$ and $\tilde W$ and properties of parameters. 
    
    Therefore, for all $t \geq 0$, $y(t)$ is in $[0,1]^{7n}$ and \eqref{eq:norm-t} holds true.
    \qed
\end{pf}\medskip

\begin{cor}\label{cor:cor1}
    Consider the adoption-opinion model \eqref{eq:vector_model} under Assumption \ref{ass:ass1}, with initial condition $y(0)$ in $[0,1]^{7n}$ satisfying \eqref{eq:norm}. Then the following hold:
    \begin{itemize}
        \item[(i)] if $x^{[k]}(0) >\boldsymbol{0}$, then $x^{[k]}(t) > \boldsymbol{0}$ for all $t \geq0$, $k \in \{1,2\}$.
        \item[(ii)] $s(t+1) \leq s(t)$ for all $t \geq0$.
    \end{itemize}
\end{cor}
\begin{pf}
    (i) From the opinion dynamics in \eqref{eq:dyn_x} and Assumption \ref{ass:ass1}, we get 
    $ x^{[k]}(t+1) \geq (I-\Lambda^{[k]}-\Xi^{[k]})x(0) > \boldsymbol{0}$ for all $k \in \{1,2\}$.
    
    (ii) From the dynamics in \eqref{eq:dyn_s}, Assumption \ref{ass:ass1} and Proposition \ref{prop:invariant}, we get that $s(t)$ is non-increasing. \qed
\end{pf}


\section{Equilibrium points and stability analysis}\label{sec:stability}
In this section, we analyze the equilibrium points of the adoption–opinion model \eqref{eq:vector_model} and their stability. We show that, under Assumption~\ref{ass:ass1}, every equilibrium is either an \emph{adoption-free} equilibrium, in which no individuals have adopted and all communities remain susceptible, or an \emph{adoption-diffused} equilibrium, characterized by a positive fraction of adopters and dissatisfied individuals across communities. These results imply that neither partial adoption (i.e., some communities adopting while others do not) nor monopolistic adoption of a single technology can occur.

\subsection{Adoption-free equilibrium}
\begin{prop}\label{prop:ade}
    Consider the adoption-opinion model \eqref{eq:vector_model} under Assumption \ref{ass:ass1}, with initial condition $y(0)$ in $[0,1]^{7n}$ satisfying \eqref{eq:norm}.
    The system admits an adoption-free equilibrium 
    \be \label{eq:y-ade}y_e:= (\boldsymbol{1},\boldsymbol{0}, \boldsymbol{0}, \boldsymbol{0}, \boldsymbol{0},{x_e^{[1]}},{x_e^{[2]}}),\ee
    where 
    \be \label{eq:x-ade}
    {x_e^{[k]}}= (I-\Lambda^{[k]} \tilde{W})^{-1} ((I- \Lambda^{[k]} - \Xi^{[k]}) x^{[k]}(0)),\mspace{7mu} k \in \{1,2\}. \ee
\end{prop}
\begin{pf}
    It is straightforward to verify that $y_e$ satisfies the firs three adoption equations in \eqref{eq:vector_model} at equilibrium. For the opinion dynamics, substituting \eqref{eq:x-ade} satisfies the equality. Note that $x_e^{[k]}$ in \eqref{eq:x-ade} is well-defined since Assumption \ref{ass:ass1} ensures $\rho(\Lambda^{[k]} \tilde{W}) < 1$, implying $I-\Lambda^{[k]} \tilde{W}$ invertible. \qed 
\end{pf}\medskip

\begin{prop}\label{prop:ade-unstable}
    Consider the adoption-opinion model \eqref{eq:vector_model} under Assumption \ref{ass:ass1}, with initial condition $y(0)$ in $[0,1]^{7n}$ satisfying \eqref{eq:norm}.
    Then the adoption-free equilibrium $y_e$ in \eqref{eq:y-ade} is not asymptotically stable.
\end{prop}
\begin{pf}
    Fix any $\varepsilon \in (0,1)$. Consider the the adoption-opinion model \eqref{eq:vector_model} with initial condition
    \[
    \begin{aligned}
        &s(0) = (1-\varepsilon)\,\boldsymbol{1},\\
        &a^{[1]}(0) = \varepsilon\,\boldsymbol{1},\qquad a^{[2]}(0) = \boldsymbol{0},\\
        &d^{[1]}(0) = \boldsymbol{0},\qquad d^{[2]}(0) = \boldsymbol{0},\\
        &x^{[k]}(0) = x_e^{[k]},\qquad k \in \{1,2\}.
    \end{aligned}
    \]
By construction,
    $$\|y(0)-y_e\|_\infty
    =\max\{\|s(0)-\boldsymbol{1}\|_\infty,\|a^{[1]}(0)-\boldsymbol{0}\|_\infty\}
    =\varepsilon.$$

    From Corollary~\ref{cor:cor1}(ii), $s(t)$ is monotonically nonincreasing, therefore, for any $t \ge 0$,
   $$s(t) \le s(0) = (1-\varepsilon)\,\boldsymbol{1}.$$
Thus,
    $$ \|s(t)-\boldsymbol{1}\|_\infty \ge \varepsilon
        \qquad \forall t\ge 0.$$

    Since $y_e$ satisfies $s_e = \boldsymbol{1}$ and all adoption-related components are zero, the above inequality implies
    \[
        \|y(t)-y_e\|_\infty 
        \;\ge\; \|s(t)-\boldsymbol{1}\|_\infty 
        \;\ge\; \varepsilon, 
        \qquad \forall\,t\ge 0.
    \]

    Hence the trajectory starting at $y(0)$ does not converge to $y_e$ and does not remain arbitrarily close to it. Since $\varepsilon>0$ was arbitrary, it follows that $y_e$ is not asymptotically stable. \qed
\end{pf}
Proposition 3 shows that the instability is structural, i.e., any arbitrarily small amount of adoption irreversibly decreases the susceptible fraction, which cannot return to~1.

\subsection{Adoption-diffused equilibrium}
The next result characterizes the adoption-diffused equilibrium, where a positive fraction of adopters is present in the network.
\begin{thm}
    Consider the adoption-opinion model \eqref{eq:vector_model} under Assumption \ref{ass:ass1}, with initial condition $y(0)$ in $[0,1]^{7n}$ satisfying \eqref{eq:norm}.
    The system admits an unique adoption-diffused equilibrium $y^*$ with ${a^{[k]}}^*, {d^{[k]}}^* >\boldsymbol{0}$ for all $k \in \{1,2\}$, $s^*=\boldsymbol{0}$ and
    \be \label{eq:x-eq}
    {x^{[k]}}^*= (I-\Lambda^{[k]} \tilde{W})^{-1} ((I- \Lambda^{[k]} - \Xi^{[k]}) x^{[k]}(0) + \Xi^{[k]}  W {a^{[k]}}^*). \ee Moreover,
    \be \label{eq:adopts_ratio} {a_i^{[2]}}^* = \frac{\delta^{[1]}_i }{\delta^{[2]}_i}{a_i^{[1]}}^* \quad \forall i.\ee
\end{thm}
\begin{pf}
    The proof proceeds in two main steps. 
    First, we establish the \emph{existence} of an adoption-diffused equilibrium by reducing the equilibrium conditions to a fixed-point equation for the adoption vector $a^{[1]*}$. 
    Second, we prove the \emph{uniqueness} of this equilibrium by exploiting the monotonicity of the associated operator and showing that the presence of two distinct fixed points would lead to a contradiction.

    \emph{Existence.} Substituting $s_i^*=0$ for all $i$ in \eqref{eq:dyn_a}-\eqref{eq:dyn_d} at equilibrium, we obtain the following relations
    \begin{align}\label{eq:equilibrium}
        {a_i^{[k]}}^* = \frac{\gamma^{[k]}_i {x^{[k]}_i}^* {d_i^{[\ell]}}^*}{\delta_i^{[k]}} = \frac{\gamma^{[\ell]}_i {x^{[\ell]}}^*_i {d_i^{[k]}}^*}{\delta_i^{[k]}}, 
    \end{align}
    for all $k,\ell \in\{1,2\}$ and $\ell \neq k$. Note that $x_i^{[k]} > 0$ for all $i$, since otherwise \eqref{eq:equilibrium} would imply $a_i^{[k]} = d_i^{[k]} = 0$, which contradicts the assumption $s_i^* = 0$ and Proposition \ref{prop:invariant}.
    We then get \eqref{eq:adopts_ratio}
    and 
    \be \label{eq:diss} {d_i^{[1]}}^* =\frac{\delta_i^{[1]} {a_i^{[1]}}^*}{\gamma^{[2]}_i {x^{[2]}_i}^*}, \quad {d_i^{[2]}}^* =\frac{\delta_i^{[1]} {a_i^{[1]}}^*}{\gamma^{[1]}_i {x^{[1]}_i}^*}.\quad \ee 
    Substituting \eqref{eq:adopts_ratio} and \eqref{eq:diss} into \eqref{eq:norm-t} at the equilibrium, we obtain the following fixed point equation in ${a_i^{[1]}}^*$:
    \be  {a_i^{[1]}}^* = 1- \frac{\delta_i^{[1]}}{\delta_i^{[2]}}{a_i^{[1]}}^*-\delta_i^{[1]}{a_i^{[1]}}^* \bigg( \frac{1}{\gamma_i^{[1]} {x_i^{[1]}}^*} + \frac{1}{\gamma_i^{[2]}{x_i^{[2]}}^*}\bigg),\ee
    where the opinion vectors ${x_i^{[1]}}^*$ and ${x_i^{[2]}}^*$ can both be expressed in terms of ${a^{[1]}_i}^*$ using \eqref{eq:x-eq} and \eqref{eq:adopts_ratio}. Therefore, we consider the following fixed-point equation in the variable $a$, 
    \be \label{eq:fixedpoint} a_i = T_i(a) := 1- \frac{\delta_i^{[1]}}{\delta_i^{[2]}}a_i-\delta_i^{[1]}a_i \bigg( \frac{1}{\gamma_i^{[1]} {x_i^{[1]}}^*\mspace{-5mu}(a)} + \frac{1}{\gamma_i^{[2]}{x_i^{[2]}}^*\mspace{-5mu}(a)}\bigg),\ee
    for all $i$.
    The vector map $T:[0,1]^n \to \mathbb{R}^n$ is continuous on the compact set $[0,1]^n$. We first show that $T$ admits at least a fixed point, which corresponds to an equilibrium for the adoption-opinion model. 
    
    First, notice that $T(\boldsymbol{0})=\boldsymbol{1}>\boldsymbol{0}$.
    We now show the existence of a vector $u\in(0,1]^n$ such that $T(u)$ will be in $[u,\boldsymbol{1}]$. 
    Note that for any $a\in[0,1]^n$ we have the uniform bounds $ {x_i^{[k]}}^* \geq x_{e,i}^{[k]}$, where $x_e^{[k]}$ is defined in \eqref{eq:x-ade}. 
    Hence, the term
    $$
    \phi_i(a) := 
    \delta_i^{[1]} \bigg( \frac{1}{\gamma_i^{[1]} {x_i^{[1]}}^*\mspace{-5mu}(a)} + \frac{1}{\gamma_i^{[2]}{x_i^{[2]}}^*\mspace{-5mu}(a)}\bigg)
    $$
    is bounded above by the finite constant
    $$
    \underline\phi_i := 
    \delta_i^{[1]} \bigg( \frac{1}{\gamma_i^{[1]}x_{e,i}^{[1]}} + \frac{1}{\gamma_i^{[2]}x_{e,i}^{[2]}}\bigg). \
    $$
    Thus, for every $a\in[0,1]^n$,
    $$
    T_i(a) \geq 1 - \frac{\delta_i^{[1]}}{\delta_i^{[2]}}a_i- \underline \phi_i a_i.
    $$
    Choose $u\in[0,1]^n$ with components
    $$
    u_i := \max\Big\{\,0,\ 1-\frac{\delta_i^{[1]}}{\delta_i^{[2]}}-\underline\phi_i\,\Big\}.
    $$
    With this choice we obtain, for every $i$ and every $a\in[u,\boldsymbol{1}]$,
    $$
    T_i(a) \geq 1 - \frac{\delta_i^{[1]}}{\delta_i^{[2]}}a_i- \underline \phi_i a_i \geq 1 - \frac{\delta_i^{[1]}}{\delta_i^{[2]}}-\underline\phi_i \geq u_i,
    $$
    hence $T([u,\boldsymbol{1}])\subseteq[u,\boldsymbol{1}]$. The set $[u,\boldsymbol{1}]$ is convex, compact and nonempty, and $T:[u,\boldsymbol{1}]\to [u,\boldsymbol{1}]$ is continuous. By Brouwer's fixed point theorem there exists $a\in [u,\boldsymbol{1}]$ such that $T(a) =a$.
By construction $a^*\neq \boldsymbol{0}$, since $T(\boldsymbol{0})=\boldsymbol{1}\neq \boldsymbol{0}$, so ${a^{[1]}}^*: = a^* >\boldsymbol{0}$. From \eqref{eq:adopts_ratio}, \eqref{eq:diss} and \eqref{eq:x-eq}, we obtain a full equilibrium $y^*$ with $s^*=\boldsymbol{0}$ and ${a^{[k]}}^*, {d^{[k]}}^* >\boldsymbol{0}$ for $k=1,2$.

    
    \emph{Uniqueness.} 
    Note that proving uniqueness reduces to showing that the operator $T$ admits a unique fixed point. 
    We analyze the Jacobian matrix of $T(a)$. 
    For $i\neq j$, the off-diagonal derivative is
    \begin{align}
    	\frac{\partial T_i}{\partial a_j} =& 
         \delta_i^{[1]} a_i \bigg( \frac{\partial {x_i^{[1]}}^*\mspace{-5mu}(a)/\partial a_j}{\gamma_i^{[1]} \big({x_i^{[1]}}^*\mspace{-5mu}(a)\big)^2} + \frac{\partial {x_i^{[2]}}^*\mspace{-5mu}(a)/\partial a_j}{\gamma_i^{[2]} \big({x_i^{[2]}}^*\mspace{-5mu}(a)\big)^2} \bigg)
        \label{eq:partial-T}
    \end{align}
    Since $\tilde W$ is row-stochastic and $\Lambda^{[k]}$ have entries less than $1$ for each technology $k$, we have that
    \be\label{eq:partial-x} \frac{\partial {x^{[k]}}^*\mspace{-5mu}(a)}{\partial a} = (I-\Lambda^{[k]} \tilde W)^{-1} \Xi^{[k]}  W \ge 0 .\ee
    Therefore, we obtain that 
    $$ \frac{\partial T_i}{\partial a_j} \geq 0, \quad \forall i  \neq j,$$
    so the Jacobian matrix of $T(a)$ is Metzler, i.e., has nonnegative off-diagonal entries.  Therefore, $T(a)$ is monotone (see \citep{hirsch2006monotone}), which means that if $a\geq b$ and $a\neq b$, this implies $T(a)> T(b)$.
    
	Observe now that for all $\alpha > 1$ and $a \in [u, 1]^n$ it holds that 
	\begin{align}
        {x^{[k]}}^*\mspace{-7mu}(\alpha a) &= \alpha {x^{[k]}}^*\mspace{-7mu}(a)- (\alpha-1) (I-\Lambda^{[k]}\tilde W)^{-1} (I-\Lambda^{[k]}-\Xi^{[k]})x^{[k]}(0) \nonumber \\
        &< \alpha x^{[k]}(a),
	\end{align}
	where the inequality follows from Proposition \ref{prop:ade}.
    Note that, in particular, for any $\alpha >1$, the following hold
    \begin{align*}
        T_i(\alpha a) &= 1-\alpha \frac{\delta_i^{[1]}}{\delta_i^{[2]}}a_i- \alpha \delta_i^{[1]}a_i \bigg( \frac{1}{\gamma_i^{[1]} {x_i^{[1]}}^*\mspace{-5mu}(\alpha a)} + \frac{1}{\gamma_i^{[2]}{x_i^{[2]}}^*\mspace{-5mu}(\alpha a)}\bigg) \\
        &<  1-\alpha \frac{\delta_i^{[1]}}{\delta_i^{[2]}}a_i- \delta_i^{[1]}a_i \bigg( \frac{1}{\gamma_i^{[1]} {x_i^{[1]}}^*\mspace{-5mu}( a)} + \frac{1}{\gamma_i^{[2]}{x_i^{[2]}}^*\mspace{-5mu}( a)}\bigg) \\
        &= \alpha T(a) +( 1-\alpha )\bigg(1-   \delta_i^{[1]}a_i \Big( \frac{1}{\gamma_i^{[1]} {x_i^{[1]}}^*\mspace{-5mu}( a)} + \frac{1}{\gamma_i^{[2]}{x_i^{[2]}}^*\mspace{-5mu}( a)}\Big) \bigg).
    \end{align*}
    Since $T(a) \in [u,1]$, the term  
    $$1-   \delta_i^{[1]}a_i \Big( \frac{1}{\gamma_i^{[1]} {x_i^{[1]}}^*\mspace{-5mu}( a)} + \frac{1}{\gamma_i^{[2]}{x_i^{[2]}}^*\mspace{-5mu}( a)}\Big) $$
    is positive and considering $\alpha >1$, we get 
    $$ \nu:= ( 1-\alpha )\Big(\frac{1}{a_i}- \delta_i^{[1]} \bigg( \frac{1}{\gamma_i^{[1]} {x_i^{[1]}}^*\mspace{-5mu}( a)} + \frac{1}{\gamma_i^{[2]}{x_i^{[2]}}^*\mspace{-5mu}( a)}\bigg) \Big) <0.$$
	Suppose now that the map $T$ admits two distinct fixed points in $[u,\boldsymbol{1}]$, so that $a' = T(a')$, $a'' = T(a'')$ and let $\alpha^* = \min_\alpha\{ \alpha a'' \geq a'\}$.
	We get that 
	\begin{align*}
	 	a'_i = T_i(a') \leq T_i(\alpha^* a'') &< \alpha^* T_i(a'') + \nu a''_i=( \alpha^*+\nu) a''_i ,
	\end{align*}
	which contradicts the definition of $\alpha^*$ as the minimum of all $\alpha$ such that $\alpha a'' \geq a'$. Therefore, we have shown that $T$ admits a unique fixed point in $[u,\boldsymbol{1}]$. Note that this implies a unique equilibrium in $d$ from \eqref{eq:equilibrium} and in $x$ from \eqref{eq:x-eq}, which leads to the uniqueness of the equilibrium for \eqref{eq:vector_model}.\qed
\end{pf}\medskip

The final result rules out the existence of \emph{partial-adoption} equilibrium point.
\begin{prop} \label{prop:no-mixed-equilibria}
    Consider the adoption-opinion model \eqref{eq:model} under Assumption \ref{ass:ass1}, with initial condition $y(0)$ in $[0,1]^{7n}$ satisfying \eqref{eq:norm} and $x^{[k]}(0)>\boldsymbol{0}$ for all $k$. Let $y^*$ be an equilibrium point. If there exists at least one node $i$ such that ${s_i}^*=0$, then ${s_j}^*=0$ for all $j$. 
\end{prop}
\begin{pf}
First note that from Assumption \ref{ass:ass1} and the fact that $x^{[k]}(0)>\boldsymbol{0}$, the equation \eqref{eq:dyn_x} at the equilibrium yields
$${x^{k}}^* = (I-\Lambda^{[k]}\tilde W)^{-1} ((I-\Lambda^{[k]}-\Xi^{[k]})x^{[k]}(0)+\Xi^{[k]} W{a^{[k]}}^* )>\boldsymbol{0},$$
for all $k \in \{1,2\}$.
Suppose by contradiction that there exists an equilibrium and a nonempty subset of nodes $J$ with ${s_i}^*=0$ for all $i\in J$, and some node $j\notin J$ with ${s_j}^*>0$.

For any technology $k$ and for each node $j \notin J$, the equilibrium equation for $s_j$ reads
$$0 = -  s_j^* \Big[\beta_j^{[1]} {x_j^{[1]}}^* \sum_{m} W_{jm} a{_m^{[1]} }^*  + \beta_j^{[2]} {x_j^{[2]}}^*  \sum_{m} W_{jm} a{_m^{[2]} }^*\Big].$$
Note that $s_j^* >0$ and $\beta_j^{[k]} {x_j^{[k]}}^*> 0$, hence
$\sum_{m} W_{jm} {a_m^{[k]}}^* = 0.$
Since $W_{jm}\ge 0$ and the graph is strongly connected, it follows that ${{a_m^{[k]}}^*=0}$ for every $m$ with $W_{jm}>0$.
Apply the same argument to each node that is reachable from $j$ in one step: all their $a^{[k]}$-components must be $0$. Iterating along directed paths (possible because the graph is strongly connected), 
we conclude that ${a_i^{k}}^*=0$ for all $i$ and for all $k$. 
The equilibrium equation for $d_i^{[k]}$ becomes
$$0 = -\gamma_i^{[\ell]} {x_i^{[\ell]}}^* {d_i^{[k]}}^* + \delta_i^{[k]} {a_i^{[k]}}^* = -\gamma_i^{[\ell]} {x_i^{[\ell]}}^* {d_i^{[k]}}^*,$$
which implies ${d_i^{[k]}}^*=0$ for all $i$ and for all $k$. 
Finally, using the normalization constraint for each node, we obtain $s_i^*=1$ for every node $i$. This contradicts the assumption that there are some nodes in $J$ with $s_i^*=0$.
Therefore no such mixed equilibrium can exist. \qed
\end{pf}

\begin{rem}[No partial-adoption equilibrium points]
    This result rules out the existence of partial-adoption equilibria when the initial opinion vector is strictly positive. In a strongly connected network, if the adoption level of even a single node vanishes at equilibrium, the coupling through the network forces all other nodes to stop adopting as well. Consequently, at equilibrium the adoption state must be either fully extinct or strictly positive on all nodes. No equilibrium exhibiting a mixed pattern of adopters and non-adopters can exist.
\end{rem}

\begin{rem}[No monopoly equilibrium points]
It is worth noting that also an equilibrium in which only one technology survives cannot occur. Indeed, suppose that at an equilibrium one technology $k$ satisfies ${a_i^{[k]}}^* = 0$ for all $i$. Then, the relations \eqref{eq:adopts_ratio} at the equilibrium lead to ${a_i^{[\ell]}}^* = 0$ for all $i$.
Thus, a configuration in which only one technology survives cannot arise at equilibrium. 
\end{rem}

\section{Numerical Simulations}\label{sec:simulations}
In this section, we present some numerical simulations of the adoption-opinion model \eqref{eq:vector_model} under different scenarios. We consider a population of $n=50$ communities. Model parameters are randomly generated, and the initial conditions correspond to an early stage of diffusion, where only a few individuals have adopted the technologies and no community has dissatisfied members.

\begin{figure}
    \centering
    \includegraphics[width=0.7\linewidth]{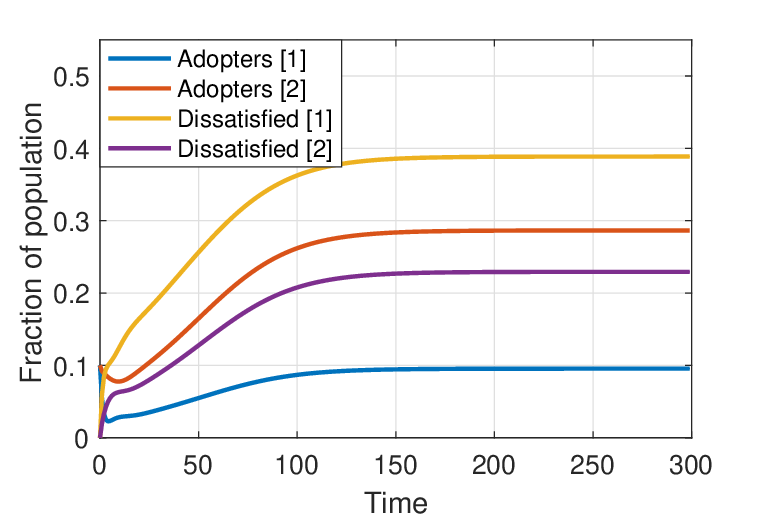}
    \caption{Numerical simulation of adoption-opinion model \eqref{eq:vector_model}.}
    \label{fig:fig1}
\end{figure}

Figure~\ref{fig:fig1} illustrates a counterintuitive effect in the adoption-opinion dynamics. Technology 1, characterized by aggressive user acquisition (high $\beta^{[1]}$) but lower perceived quality (higher dissatisfaction rate $\delta^{[1]}$), initially attracts a large fraction of users. However, due to the high dissatisfaction, many users quickly transition to the dissatisfied state, creating a pool of potential adopters for Technology 2. In contrast, Technology 2, with moderate adoption effort ($\beta_i^{[2]} < \beta_i^{[1]}$ for each community $i$) but higher perceived quality ($\delta_i^{[2]} < \delta_i^{[1]}$ for each community $i$), gradually converts these dissatisfied users into adopters. The dynamics thus reveal that intensive marketing without sufficient quality can paradoxically favor a competitor with better user experience.

\begin{figure}
    \centering
    \includegraphics[width=0.7\linewidth]{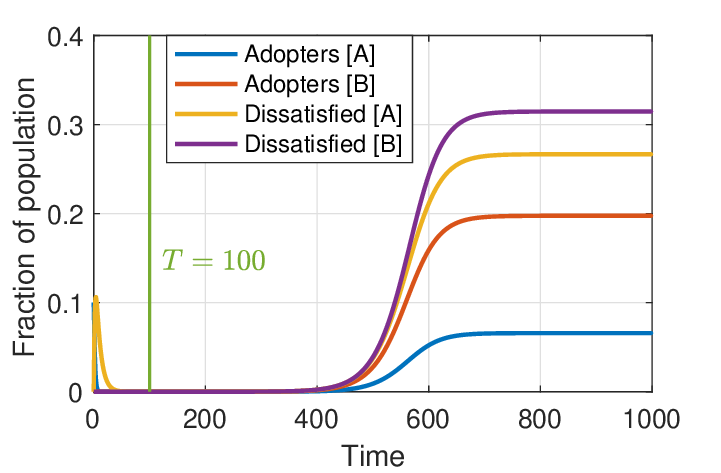}
    \caption{Numerical simulation of adoption-opinion model \eqref{eq:vector_model} where Technology $2$ enters the market after a delay $T=100$.}
    \label{fig:fig2}
\end{figure}

Figure~\ref{fig:fig2} shows the scenario in which Technology 2 enters the market with a delay $T= 100$. Despite the late entry, the qualitative behavior of the system remains similar: Technology 1 still attracts early adopters, many of whom become dissatisfied, while the higher-quality late entrant gradually gains adoption. This demonstrates that the counterintuitive effect is robust with respect to entry timing: a superior user experience eventually dominates adoption dynamics, even if the technology enters later.

From a theoretical and control perspective, these simulations illustrate an important feature of the equilibrium (as shown in \eqref{eq:equilibrium}): the fraction of adopters at the equilibrium does not depend on the adoption rates $\beta^{[k]}$ or on the opinions vector $x^{[k]}$. This implies that interventions aimed purely at increasing marketing or giving adoption incentives will have limited effect on the ultimate market shares. 
Interestingly, even symmetric campaigns that increase average opinion will disproportionately benefit the higher-quality technology, a clear example of symmetric control leading to asymmetric outcomes.
Instead, the primary control lever is the dissatisfaction rate $\delta^{[k]}$, which captures the perceived quality and user experience of the technology. By reducing dissatisfaction, through product improvement, enhanced user experience, or better post-adoption support, a policy maker can increase the fraction of adopters relative to competitors \eqref{eq:adopts_ratio}. These results suggest that, in strategic terms, focusing on quality and satisfaction is more effective than aggressive marketing in shaping long-term adoption outcomes, and that policies or interventions aimed at enhancing user experience are likely to yield higher and more sustainable market penetration.

\section{Conclusion}\label{sec:conclusion}
This work introduces a two-layer model coupling opinion formation and technology adoption in the presence of two competing technologies. We demonstrate the existence and uniqueness of the adoption–diffused equilibrium in which both technologies coexist. Our analytical findings show that coexistence is intrinsic to this coupled process: neither monopoly nor partial-adoption states can emerge in the long run. Numerical experiments highlight that quality and user experience are the key factors determining long-term dominance, while symmetric interventions on opinions or adoption rates can lead to markedly asymmetric outcomes that favor the superior technology.
These results provide a deeper understanding of how technological competition unfolds in interconnected societies and offer guidance for designing effective innovation management strategies. Future research will focus on establishing formal convergence guarantees toward the equilibrium and extending the modeling framework to include more than two competing technologies, enabling the study of richer and more realistic competitive markets.

\bibliography{bib.bib}  

@article{BRESCHI2023103651,
    title = {Driving electric vehicles’ mass adoption: An architecture for the design of human-centric policies to meet climate and societal goals},
    journal = {Transportation Research Part A: Policy and Practice},
    volume = {171},
    pages = {103651},
    year = {2023},
    issn = {0965-8564},
    author = {Valentina Breschi and Chiara Ravazzi and Silvia Strada and Fabrizio Dabbene and Mara Tanelli},
}

@article{VILLA2024106106,
    title = {Can control aid in attaining sustainable goals? {A}n improved data-informed framework to promote shared mobility},
    journal = {Control Engineering Practice},
    volume = {153},
    pages = {106106},
    year = {2024},
    issn = {0967-0661},
    author = {Eugenia Villa and Valentina Breschi and Chiara Ravazzi and Mara Tanelli and Fabrizio Dabbene},
}

@article{Bass1969,
	author = {Bass, Frank M.},
	title = {A New Product Growth for Model Consumer Durables},
	journal = {Management Science},
	volume = {15},
	number = {5},
	pages = {215-227},
	year = {1969}
}

@article{Pastor-Satorras2015,
	author = {Pastor-Satorras, Romualdo and Castellano, Claudio and Van Mieghem, Piet and Vespignani, Alessandro},
	title = {Epidemic processes in complex networks},
	journal = {Reviews of Modern Physics},
	volume = {87},
	number = {3},
	pages = {925-979},
	year = {2015}
}

@article{Granovetter1978,
	author = {Granovetter, Mark},
	title = {Threshold Models of Collective Behavior},
	journal = {American Journal of Sociology},
	volume = {83},
	number = {6},
	pages = {1420-1443},
	year = {1978}
}

@article{Friedkin1990,
	author = {Friedkin, Noah E. and Johnsen, Eugene C.},
	title = {Social Influence and Opinions},
	journal = {Journal of Mathematical Sociology},
	volume = {15},
	number = {3-4},
	pages = {193-206},
	year = {1990}
}

@article{Kermack.McKendrick:1927,
	author = {Kermack, W. O. and McKendrick, A. G.},
	journal = {Proceedings of the {R}oyal {S}ociety of {L}ondon. Series A},
	number = {772},
	pages = {700--721},
	title = {A contribution to the mathematical theory of epidemics},
	volume = {115},
	year = {1927}}

@article{Xu2024,
	author = {Q. Xu and H. Ishii},
	title = {On a Discrete-Time Networked {SIV} Epidemic Model with Polar Opinion Dynamics},
	journal = {IEEE Transactions on Network Science and Engineering},
	volume = {11},
	number = {6},
	pages = {6636-6651},
	year = {2024}
}

@article{DeGroot1974,
	author = {M. H. DeGroot},
	title = {Reaching a consensus},
	journal = {Journal of the American Statistical Association},
	volume = {69},
	number = {345},
	pages = {118--121},
	year = {1974}
}

@article{Altafini2012,
	author = {C. Altafini},
	title = {Consensus problems on networks with antagonistic interactions},
	journal = {IEEE Transactions on Automatic Control},
	volume = {58},
	number = {4},
	pages = {935--946},
	year = {2012}
}

@article{LI20141042,
	title = {Analysis of epidemic spreading of an {SIRS} model in complex heterogeneous networks},
	journal = {Communications in Nonlinear Science and Numerical Simulation},
	volume = {19},
	number = {4},
	pages = {1042-1054},
	year = {2014},
	issn = {1007-5704},
	author = {Chun-Hsien Li and Chiung-Chiou Tsai and Suh-Yuh Yang},
}

@article{ZHANG2021126524,
	title = {Layered {SIRS} model of information spread in complex networks},
	journal = {Applied Mathematics and Computation},
	volume = {411},
	pages = {126524},
	year = {2021},
	issn = {0096-3003},
	author = {Yuexia Zhang and Dawei Pan},
}

@inproceedings{Kempe2003,
    author = {Kempe, David and Kleinberg, Jon and Tardos, \'{E}va},
    title = {Maximizing the spread of influence through a social network},
    year = {2003},
    publisher = {Association for Computing Machinery},
    address = {New York, NY, USA},
    booktitle = {Proceedings of the Ninth ACM SIGKDD International Conference on Knowledge Discovery and Data Mining},
    pages = {137–146},
    numpages = {10},
}

@article{montanari2010spread,
	title={The spread of innovations in social networks},
	author={Montanari, Andrea and Saberi, Amin},
	journal={Proceedings of the National Academy of Sciences},
	volume={107},
	number={47},
	pages={20196--20201},
	year={2010},
	publisher={National Academy of Sciences}
}

@article{young2011dynamics,
	title={The dynamics of social innovation},
	author={Young, H Peyton},
	journal={Proceedings of the National Academy of Sciences},
	volume={108},
	number={supplement\_4},
	pages={21285--21291},
	year={2011},
	publisher={National Academy of Sciences}
}

@book{Rogers2010,
	title={Diffusion of Innovations},
	author={Rogers, Everett M.},
	year={2010},
	publisher={Simon and Schuster},
	address={New York}
}

@article{bettencourt2006power,
	title={The power of a good idea: Quantitative modeling of the spread of ideas from epidemiological models},
	author={Bettencourt, Lu{\'\i}s MA and Cintr{\'o}n-Arias, Ariel and Kaiser, David I and Castillo-Ch{\'a}vez, Carlos},
	journal={Physica A: Statistical Mechanics and its Applications},
	volume={364},
	pages={513--536},
	year={2006},
	publisher={Elsevier}
}

@article{goffman1966mathematical,
	title={Mathematical approach to the spread of scientific ideas - {T}he history of mast cell research},
	author={Goffman, William},
	journal={Nature},
	volume={212},
	number={5061},
	pages={449--452},
	year={1966},
	publisher={Nature Publishing Group UK London}
}

@article{granell2013dynamical,
	title={Dynamical interplay between awareness and epidemic spreading in multiplex networks},
	author={Granell, Clara and G{\'o}mez, Sergio and Arenas, Alex},
	journal={Physical Review Letters},
	volume={111},
	number={12},
	pages={128701},
	year={2013},
	publisher={APS}
}

@article{wang2019impact,
	title={The impact of awareness diffusion on {SIR}-like epidemics in multiplex networks},
	author={Wang, Zhishuang and Guo, Quantong and Sun, Shiwen and Xia, Chengyi},
	journal={Applied Mathematics and Computation},
	volume={349},
	pages={134--147},
	year={2019},
	publisher={Elsevier}
}

@inproceedings{lin2021discrete,
	title={On a discrete-time network {SIS} model with opinion dynamics},
	author={Lin, Yixuan and Xuan, Weihao and Ren, Ruijie and Liu, Ji},
	booktitle={Proceedings of the 2021 60th IEEE Conference on Decision and Control (CDC 2021)},
	pages={2098--2103},
	year={2021},
}

@inproceedings{she2021peak,
	title={Peak infection time for a networked {SIR} epidemic with opinion dynamics},
	author={She, Baike and Leung, Humphrey CH and Sundaram, Shreyas and Par{\'e}, Philip E},
	booktitle={Proceedings of the 2021 60th IEEE Conference on Decision and Control (CDC 2021)},
	pages={2104--2109},
	year={2021},
}

@inproceedings{bizyaeva2024active,
	title={Active risk aversion in {SIS} epidemics on networks},
	author={Bizyaeva, Anastasia and Arango, Marcela Ordorica and Zhou, Yunxiu and Levin, Simon and Leonard, Naomi Ehrich},
	booktitle={IEEE 2024 American Control Conference (ACC)},
	pages={4428--4433},
	year={2024},
}

@book{centola2018behavior,
	title={How Behavior Spreads: The Science of Complex Contagions},
	author={Centola, Damon},
	volume={3},
	year={2018},
	publisher={Princeton University Press, Princeton, NJ}
}

@article{alutto2025predictive,
  title={Predictive Control Strategies for Sustaining Innovation Adoption on Multilayer Social Networks},
  author={Alutto, Martina and Xu, Qiulin and Dabbene, Fabrizio and Ishii, Hideaki and Ravazzi, Chiara},
  journal={arXiv preprint arXiv:2509.01457},
  year={2025}
}

@article{alutto2025modeling,
  title={Modeling and Control of Sustainable Transitions through Opinion-Behavior Coupling in Heterogeneous Networks},
  author={Alutto, Martina and Bellotti, Sofia and Dabbene, Fabrizio and Ravazzi, Chiara},
  journal={arXiv preprint arXiv:2511.11053},
  year={2025}
}

@article{arthur1989competing,
  title={Competing technologies, increasing returns, and lock-in by historical events},
  author={Arthur, W Brian},
  journal={The Economic Journal},
  volume={99},
  number={394},
  pages={116--131},
  year={1989},
  publisher={Oxford University Press Oxford, UK}
}

@article{david1985clio,
  title={Clio and the Economics of {QWERTY}},
  author={David, Paul A},
  journal={The American Economic Review},
  volume={75},
  number={2},
  pages={332--337},
  year={1985},
  publisher={JSTOR}
}

@article{witt1997lock,
  title={“{L}ock-in” vs.“critical masses” - {I}ndustrial change under network externalities},
  author={Witt, Ulrich},
  journal={International Journal of Industrial Organization},
  volume={15},
  number={6},
  pages={753--773},
  year={1997},
  publisher={Elsevier}
}

@article{min2018competing,
  title={Competing contagion processes: Complex contagion triggered by simple contagion},
  author={Min, Byungjoon and San Miguel, Maxi},
  journal={Scientific Reports},
  volume={8},
  number={1},
  pages={10422},
  year={2018},
  publisher={Nature Publishing Group UK London}
}

@article{hirsch2006monotone,
  title={Monotone dynamical systems},
  author={Hirsch, Morris W and Smith, Hal},
  journal={Handbook of differential equations: ordinary differential equations},
  volume={2},
  pages={239--357},
  year={2006},
  publisher={Elsevier}
}

@article{liu2019analysis, 
title={Analysis and control of a continuous-time bi-virus model}, 
author={Liu, Ji and Par{\'e}, Philip E and Nedi{\'c}, Angelia and Tang, Choon Yik and Beck, Carolyn L and Ba{\c{s}}ar, Tamer}, 
journal={IEEE Transactions on Automatic Control}, 
volume={64}, 
number={12}, 
pages={4891--4906}, 
year={2019}, 
}

@article{gracy2025modeling, 
title={Modeling and analysis of a coupled {SIS} bi-virus model}, 
author={Gracy, Sebin and Par{\'e}, Philip E and Liu, Ji and Sandberg, Henrik and Beck, Carolyn L and Johansson, Karl Henrik and Ba{\c{s}}ar, Tamer}, 
journal={Automatica}, 
volume={171}, 
pages={111937}, 
year={2025}, 
publisher={Elsevier} }
\end{document}